\documentclass[preprint,12pt]{elsarticle}
\usepackage{a4wide}
\usepackage{color}

\usepackage[ansinew]{inputenc}
\usepackage{booktabs}
\usepackage{amsmath}
\usepackage{amsthm}
\usepackage{amssymb}
\usepackage{amsfonts}
\usepackage{graphicx}
\usepackage{caption}
\usepackage{subcaption}
\usepackage[english]{babel}
\usepackage{floatrow}
\usepackage{float}
\usepackage{epsfig}
\usepackage{epstopdf}
\usepackage{grffile}
\usepackage{blkarray}
\usepackage {tikz}
\usepackage{multirow} 
\usepackage{rotating} 
\usepackage{pdflscape}
\usepackage{afterpage}
\usetikzlibrary {positioning}
\definecolor {processblue}{cmyk}{0.96,0,0,0}

\allowdisplaybreaks[1]

\journal{Journal of Computational and Applied Mathematics}

\begin{document}
\begin{frontmatter}



\title{High-order compact finite difference scheme for option pricing
  in stochastic volatility jump models}


\author[label1]{Bertram D{\"u}ring\corref{cor1}} 
\cortext[cor1]{Corresponding author}
\address[label1]{Department of Mathematics, University of Sussex, Pevensey II, Brighton, BN1 9QH, United Kingdom.}
\ead{bd80@sussex.ac.uk}

\author[label1]{Alexander Pitkin}
\ead{A.H.Pitkin@sussex.ac.uk}

\begin{abstract}
We derive a new high-order compact finite difference scheme for option
pricing in stochastic volatility jump models, e.g.\ in Bates model. In
such models the option price is determined as the solution of a
partial integro-differential equation.
The scheme is fourth order accurate in space and second order accurate
in time. 
Numerical experiments for
the European option pricing problem are presented. We validate the stability of
the scheme numerically and compare its 
performance to standard finite difference and finite element methods.  
The new scheme
outperforms a standard discretisation based on a second-order central
finite difference approximation in all our experiments. At the same time, it is very efficient, requiring only one initial $LU$-factorisation of a
sparse matrix to perform the option price valuation. Compared to
finite element approaches, it is very parsimonious in terms of memory requirements
and computational effort, since it achieves high-order convergence
without requiring additional unknowns, unlike finite element methods with higher
polynomial order basis functions. The new high-order compact scheme can also be
useful to upgrade existing implementations based on
standard finite differences in a straightforward manner to obtain a highly efficient option
pricing code.
\end{abstract}


\end{frontmatter}

\section{Introduction}
\noindent The classical model for pricing financial options is the
model of Black and Scholes \cite{BlaSch73} who consider that the underlying follows a
geometric Brownian motion with constant volatility. This allows for the derivation of
simple, closed-form option price formulae, however it is
unable to explain commonly observed features of option market
prices, like the implied volatility smile (or smirk) and excess and random volatility. 
A wide range of option pricing models have been proposed in the
literature to alleviate such shortcomings. Many of the most widely used models share
one of the following two features: (i) introduction of further risk
factors, very often stochastic volatility \cite{Lewis00}, most famously in the Heston
stochastic volatility model \cite{Hes93}; (ii) jumps in the underlying
stochastic processes, e.g.\ as already introduced by Merton \cite{Mer76}. 
In 1996, Bates \cite{Bates} proposed to combine both features in one
model, now commonly referred to as the Bates or Stochastic Volatility
Jump (SVJ) model. In this model the option price
is given as the solution of a partial integro-differential equation (PIDE),
see e.g.\ \cite{Cont}. It is able to capture the typical features of
market option prices, allowing for improved flexibility introduced by
stochastic volatility and at the same time being able to fit short-dated skews by the
incorporation of jumps in the underlying's process.
It now takes the position of a quasi market standard in option pricing
applications.

For some option pricing models closed-form solutions are available for vanilla payoffs
(see e.g.\ \cite{DPS00}) or at least approximate analytic expressions, see
e.g.\ \cite{BeGoMi10} and the literature cited therein. In general, however, one has to rely on numerical methods for pricing options.

For numerical methods for option pricing models with a single risk factor, leading to
partial differential equations in one spatial dimension, e.g.\ variants of the
the Black-Scholes model, there is a large mathematical
literature, with many relying on standard finite
difference methods (see e.g.\ \cite{TavRan00} and the references
therein). 
For one-dimensional models with jump-diffusion we refer to
\cite{Cont,Duffie05,Briani,Sachs,Salmi}.

For option pricing models with more than one risk factor,
e.g.\ in stochastic volatility models, which involve solving
partial differential equations in two or more spatial
dimensions, there are fewer works, e.g.\ \cite{IkoToi07} where different efficient
methods for solving the American option pricing problem for the Heston
model are proposed. Other approaches include finite element-finite
volume \cite{ZvFoVe98}, multigrid \cite{ClaPar99}, sparse wavelet
\cite{HiMaSc05}, FFT-based \cite{Oosterlee}, spectral
\cite{ZhuKop10}, hybrid tree-finite difference \cite{Briani15} methods and operator splitting techniques \cite{HouFou10, DuFoRi13,DuMi17,HeHeEhGu17,DuHeMi17}. 

For problems which additionally include jumps in the
underlying's process, and require the solution of PIDE in two or more
spatial dimensions, there are even fewer works. We mention
\cite{Salmi14,vonSydow} who propose an implicit-explicit time
discretisation in combination with a standard, second-order finite
difference discretisation in space and \cite{FaCoJo14} who discuss and
analyse an explicit discretisation. A method
of lines algorithm for pricing American options under the Bates model
is presented in \cite{Chiarella09}.
An alternative approach is
discussed in \cite{Briani16}, where the authors combine tree methods and 
finite differences in a hybrid scheme for the Bates model with
stochastic interest rates.

More recently, high-order finite difference schemes (fourth order in space)
have been proposed for solving partial differential equations
arising from stochastic volatility models. In \cite{DuFo12} a high-order
compact finite difference scheme for option pricing in the Heston
model is derived. This approach is extended to non-uniform grids in
\cite{DuFoHe14}, and to multiple space dimensions in \cite{DuHe15}.

High-order compact schemes have in the literature originally
been proposed for the numerical approximation of solutions to rather specific problems, as
the Poisson or the heat equation. Only gradually over the last two decades has
progress been made to extend this approach to more complex, and
multi-dimensional or nonlinear, problems. The derivation of high-order
compact schemes is algebraically demanding and hence these schemes are often tailored to rather
specific problems.

The originality of the present work consists in proposing a new {\em implicit-explicit high-order compact finite difference scheme\/} for option
pricing in Bates model.  Up to the knowledge of the authors it presents the
first high-order scheme for this highly popular option pricing model.
It combines a ---suitably adapted--- version of the high-order 
compact scheme from \cite{DuFo12}  with an explicit treatment of the
integral term which matches the high-order, inspired by the work of
Salmi {\em et al.} \cite{Salmi14}. 
The new compact scheme is fourth order accurate in space
and second order accurate in time. 
We validate the stability of
the scheme numerically and compare its
performance to both standard finite difference methods and finite
element approaches. The new scheme
outperforms a standard discretisation based on a second-order central
finite difference approximation.
Compared to the finite element approach, it is very parsimonious in
terms of memory requirements and computational effort, since it
achieves high-order convergence without requiring additional unknowns
---unlike finite element methods with higher polynomial order.
At the same time, the new high-order compact scheme is very efficient, requiring only one initial $LU$-factorisation of a
sparse matrix to perform the option price valuation. 
It can also be
useful to upgrade existing implementations based on
standard finite differences in a straightforward manner to obtain a highly efficient option
pricing code. 

This article is organised as follows. In the next section we recall Bates
model for option pricing and the related
partial integro-differential equation. Section~\ref{sec:transf} 
is devoted to a variable transformation for the problem.
The new scheme is derived in Section~\ref{sec:impexp}.
The smoothing of the initial condition and the discretisation of the boundary conditions are discussed in
Section~\ref{sec:initboundcond}. In Section~\ref{sec:FEM} we state the
finite element formulation which we use for the numerical comparison
experiments. In Section~\ref{sec:num} we present numerical convergence
and stability results, investigate and compare the efficiency of
the scheme to other methods, and study its
hedging performance. Section~\ref{sec:conc} concludes.

\section{Bates Model}

\noindent We recall the Bates model \cite{Bates} which we focus our paper on. The Bates model is a stochastic volatility model which allows for jumps in returns. Within this model the behaviour of the asset value, \textit{S}, and its variance, \( \sigma \), is described by the coupled stochastic differential equations,
\begin{align*}
dS(t) &= \mu_B S(t) dt + \sqrt{\sigma (t)} S(t) dW_1(t) + S(t) dJ,\\
d\sigma(t) &= \kappa(\theta - \sigma(t)) + v\sqrt{\sigma (t)} dW_2(t),
\end{align*}
for $ 0\leqslant t \leqslant T $ and with $ S(0), \sigma(0) > 0  $. Here, $ \mu_B = r - \lambda\xi_B $ is the drift rate, where $r\geqslant0$ is the risk-free interest rate. The jump process $ J $ is a compound Poisson process with intensity $ \lambda\geqslant0$ and $J+1$ has a log-normal distribution $p(\tilde{y})$ with the mean in $\log (\tilde{y})$ being $\gamma$ and the variance in $\log (\tilde{y})$ being $v^2$, i.e.\ the probability density function is given by
 \[p(\tilde{y})=\frac{1}{\sqrt{2\pi}\tilde{y}v}e^{-\frac{(\log\tilde{y}-\gamma)^2}{2v^2}}.\]
The parameter $\xi_B$ is defined by $\xi_B=e^{\gamma+\frac{v^2}{2}}-1$. The variance has mean level $\theta$, $\kappa$ is the rate of reversion back to mean level of $\sigma$ and $v$ is the volatility of the variance $\sigma$. The two Wiener processes $W_1$ and $W_2$ have correlation $\rho$.

By standard derivative pricing arguments for the Bates model, we
obtain the partial integro-differential equation
\begin{multline}
\label{eq:Bates}
\frac{\partial V}{\partial t} +
  \frac{1}{2}S^2\sigma\frac{\partial^2 V}{\partial S^2}+\rho v\sigma
  S\frac{\partial^2 V}{\partial S \partial \sigma}
  +\frac{1}{2}v^2\sigma \frac{\partial^2 V}{\partial \sigma^2} +
  (r-\lambda\xi_B)S\frac{\partial V}{\partial S} + \kappa(\theta -
  \sigma) \frac{\partial V}{\partial \sigma} - (r+\lambda)V \\+
  \lambda \int_0^{+\infty} \! V(S\tilde{y},v,t)p(\tilde{y}) \,
  \mathrm{d}\tilde{y} = L_D V+L_I V,
\end{multline}
which has to be solved for $S,\sigma > 0$, $0 \leq t < T $ and subject
to a suitable final condition, e.g.\ $V(S,\sigma,T) = \max(K-S,0), $
in the case of a European put option, with $K$ denoting the strike price. For clarity the operators $L_D V$ and $L_I V$ are defined as the differential part (including the term $-(r+\lambda)V$) and the integral part, respectively.

\section{Transformation of the equation}
\label{sec:transf} 

\noindent Using the transformation of variables
$$ x=\log S , \quad \tau = T-t , \quad   y=\frac{\sigma}{v} \quad \text{and} \quad u= \exp (r+\lambda)V,  $$
we obtain
\begin{multline} 
\label{eq:Ptransf}
u_\tau = \frac{1}{2}vy\left(\frac{\partial^2 u
    }{\partial x^2}+\frac{\partial^2 u}{\partial y^2}\right)+\rho
  vy\frac{\partial^2 u}{\partial x \partial y}
  -\left(\frac{1}{2}vy-r+\lambda\xi_B\right)\frac{\partial u}{\partial
    x} + \kappa \frac{(\theta - vy) }{v}\frac{\partial u}{\partial y}
  +\exp (r+\lambda)L_I V,
\end{multline}
which is now posed on $ \mathbb{R} \times \mathbb{R} ^+ \times (0,T), $ with
 \[ L_I V = \lambda \int_0^\infty \! V(S\tilde{y},v,t)p(\tilde{y}) \, \mathrm{d}\tilde{y}. \] 
Applying the same transformation to the intergral term, $L_I$,
\[ \exp (r+\lambda)L_I V = \lambda \int_0^{+\infty}  u(x \tilde y , y , \tau ) p(\tilde y) \, \mathrm{d}\tilde y . \]
Now by setting $ z=\log \tilde y,$ $\tilde u (z,y,\tau) = u(e^z,y,\tau) $ and $ \tilde  p (z) = e^z p(e^z) $ we have
\[ \exp (r+\lambda)L_I V = \lambda \int_0^{+\infty}  u(x \tilde y , y , \tau ) p(\tilde y) \, \mathrm{d}\tilde y  = \lambda \int^{+\infty}_{-\infty} \tilde u(x + z , y , \tau ) \tilde p(z) \, \mathrm{d}z.  \]
The problem is completed by the following initial and boundary conditions: 
\begin{align*}
 u(x,y ,0)&= \max (1- \exp(x),0), \quad x \in \mathbb{R}, \; y > 0, \\
 u(x,y,t) &\rightarrow 1, \quad x \rightarrow - \infty , \; y  > 0, \; t > 0, \\
 u(x,y,t) &\rightarrow 0, \quad x \rightarrow + \infty , \; y > 0, \; t > 0,  \\
u_y (x,y,t) &\rightarrow 0, \quad x \in \mathbb{R} , \; y \rightarrow \infty , \; t > 0,  \\
 u_y(x,y,t) &\rightarrow 0, \quad x \in \mathbb{R} , \;
  y  \rightarrow 0, \; t > 0. 
\end{align*}

\section{Implicit-explicit scheme}
\label{sec:impexp}

\noindent Following the idea employed by Salmi, Toivanen and von Sydow
in \cite{Salmi14,vonSydow}, we accomplish the implicit-explicit discretisation in time by means of the IMEX-CN method. This method is an adaptation of the Crank-Nicholson method, whereby an explicit treatment is added for the integral operator.
To achieve high-order convergence we adapt the high-order compact finite difference
scheme developed in \cite{DuFo12} to implicitly approximate the
differential operator, while we evaluate the integral explicitly using
the Simpson's rule to match the high-order accuracy of the high-order
compact scheme.

\subsection{High-order compact scheme for the differential operator}

\noindent Following the discretisation employed in \cite{DuFo12}, we replace
$\mathbb{R}$ by $[-R_1,R_1]$ and $\mathbb{R}^+$ by $[L_2,R_2]$ with
$R_1,R_2 >L_2>0$. We consider a uniform grid $ Z = \{ x_i \in [-R_1,R_1] : x_i = i h_1 ,\; i =-N, ...  ,N\} \times \{y_j \in [L_2,R_2] : y_j = L_2 + j h_2 ,\; j=0, ... , M\} $ consisting of $(2N+1) \times (M+1) $ grid points with $R_1 = N h_1$ , $R_2 = L_2 + M h_2 $ and with space steps $ h_1, h_2 $ and time step $ k $. Let $ u_{i,j}^n $ denote the approximate solution of (2) in $(x_i, y_j) $ at the time $t_n = n k$ and let $u^n = (u_{i,j}^n) $.    

\subsubsection{Elliptic problem}

\noindent We introduce the high-order compact discretisation for the elliptic
problem with Laplacian operator, 
\begin{multline} 
\label{eq:Lelliptic}
-\frac{1}{2}vy\left(\frac{\partial^2 u }{\partial
      x^2}+\frac{\partial^2 u}{\partial y^2}\right)-y\rho
  v\frac{\partial^2 u}{\partial x \partial y}
  -\left(r-\frac{1}{2}vy-\lambda\xi_B\right)\frac{\partial u}{\partial
    x} - \kappa \frac{(\theta - vy) }{v}\frac{\partial u}{\partial y}
  = f(x,y).
 \end{multline}
We construct a fourth-order compact finite difference scheme with a
nine-point computational stencil using the eight nearest neighbouring
points around a reference grid point $(i,j)$, following the approach
in \cite{DuFo12}.  The idea behind the derivation of the high-order compact scheme is to operate
  on the differential equations as an auxiliary relation to obtain finite difference
  approximations for high-order derivatives in the truncation error. Inclusion
  of these expressions in a central difference approximation increases
  the order of accuracy while retaining a compact computational stencil.

 Introducing a uniform grid with mesh spacing $ h=h_1 = h_2 $ in both the $x$- and $y$-directions, the standard central difference approximation to equation \eqref{eq:Lelliptic} at grid point $(i,j)$ is 
\begin{multline}
\label{eq:Lellipticdisc}
 - \frac{1}{2}vy_j \left( \delta_x^2 u_{i,j} + \delta_y^2
  u_{i,j}   \right) -\rho vy_j \delta_x\delta_y u_{i,j}  +
  \left(\frac{1}{2}vy_j-r+\lambda\xi_B\right)\delta_x u_{i,j}\\ - \kappa
  \frac{(\theta - vy_j) }{v} \delta_y u_{i,j} - \tau_{i,j}  = f(i,j) , 
\end{multline}
  where $\delta_x$ and $\delta_x^2$ ($\delta_y$ and $\delta_y^2$, respectively) denote 
  the first and second order central difference approximations with
  respect to $x$ (with respect to $y$).
The associated truncation error is given by
\begin{multline}  
\label{eq:tau}
\tau_{i,j}  = \frac {1}{24} vyh^2 \left(u_{xxxx} + u_{yyyy} \right) + \frac{1}{6} \rho vyh^2 \left( u_{xyyy} + u_{xxxy} \right) + \frac{1}{12} \left( 2r - vy - 2\lambda\xi_B \right) h^2 u_{xxx}   \\  +\frac{1}{6} \frac{\kappa (\theta - vy) }{v} h^2 u_{yyy} + \mathcal{O}(h^4).
\end{multline}
For the sake of clarity the subindices $j$ and $(i,j)$ on $y_j$ and $u_{i,j}$ (and its derivatives) are omitted from here. Differentiating \eqref{eq:Lelliptic} with respect to $x$ and $y$, respectively, yields,
\begin{align} 
\label{eq:aux1}
u_{xxx} &= -u_{xyy} -2\rho u_{xxy} +\frac{2\lambda\xi_{B} +vy -2r}{vy}
  u_{xx} - \frac{2\kappa (-vy + \theta)}{yv^2} u_{xy} - \frac{2}{vy}
  f_x , \\
 u_{yyy} &= -u_{yxx} -2\rho u_{yyx} - \frac{1}{y} u_{xx}
           +\frac{2\lambda\xi_{B} - 2\rho v +vy -2r}{vy} u_{yx}
           \nonumber\\ 
&\hspace*{4cm}-
           \frac{-2\kappa vy  + 2\kappa \theta +v^2}{v^2y} u_{yy}
           +\frac{1}{y} u_x + \frac{2\kappa}{vy} u_y - \frac{2}{vy}
  f_y . \label{eq:aux2}
\end{align}
Differentiating equations \eqref{eq:aux1} and \eqref{eq:aux2} with respect to $y$ and $x$, respectively, and adding the two expressions we obtain
\begin{multline}
\label{eq:aux3}
 u_{{{xyyy}}}+u_{{{xxxy}}} = -2\,\rho\,u_{{{\it xxyy}}}-\,{
\frac {u_{{{\it xxx}}}}{2y}}+{\frac { \left( 2\,\lambda\,\xi_{{B}}-\rho
\,v+vy-2\,r \right) u_{{{\it xxy}}}}{vy}}\\-\,{\frac { \left( -4\,
\kappa\,vy+4\,\kappa\,\theta+{v}^{2} \right) u_{{{\it xyy}}}}{2 y{v}^{2}
}}  -\,{\frac { \left( 2\,\lambda\,\xi_{{B}}-vy-2\,r \right) u_{{{
\it xx}}}}{2v{y}^{2}}} +{\frac {\kappa\, \left( vy+\theta \right) u_{{{
\it xy}}}}{{y}^{2}{v}^{2}}}+{\frac {f_{{x}}}{v{y}^{2}}} .
\end{multline}
By differentiating equation \eqref{eq:Lelliptic} twice with respect to $x$ and twice with respect to $y$ and adding the two expressions, we obtain
\begin{multline} 
\label{eq:aux4}
u_{{{ xxxx}}}+ u_{{{ yyyy}}} =
-2 \rho u_{{\it xxxy}} - 2 \rho u_{{{\it xyyy}}} -2\,u_{{{\it xxyy}}}
+ 2 \frac { \left( \kappa vy - v^2 -\kappa\,\theta \right)}{v^2 y} u_{{{\it xxy}}} 
- \frac { \left( 2r- vy-2\lambda\,\xi_{B} \right)}{vy} u_{{{\it xxx}}} \\
+ 2 \frac { \left( \kappa vy- v^2 - \kappa\,\theta \right)}{v^2 y} u_{{{\it yyy}}}
- \frac { \left( -vy+4 \rho v -2 \lambda\,\xi_{{B}}+2 r \right)}{vy} u_{{{\it xyy}}} 
+4 \frac {\kappa}{vy} u_{{{\it yy}}} + \frac{2}{y} {u_{{{\it xy}}}} - \frac{2}{vy} \left(f_{xx} + f_{yy} \right).
\end{multline}
We now substitute equations \eqref{eq:aux1}--\eqref{eq:aux4} into
\eqref{eq:tau} to yield a new expression of the error term $\tau_{i,j}$ that only consists of terms which are either $ \mathcal{O}(h^4) $ or $ \mathcal{O}(h^2) $ 
multiplied by derivatives of $u$ which can be approximated to  $
\mathcal{O}(h^2) $ within the compact stencil. Inserting this new
expression for the error term in \eqref{eq:Lellipticdisc} we obtain the
  following $\mathcal{O}(h^4)$ approximation to the partial
  differential equation (\ref{eq:Lelliptic}),
\begin{multline}
\label{eq:HOCelliptic}
-\frac{1}{24} {\frac{ 4 {h}^{2}{\lambda}{\xi_{{B}}} \left( {\lambda}{\xi_{{B}}} +\rho v - 2 \right)  
+vy_j \left( vy_j - 2 \kappa - 2 r \right) - 2 \left(r \rho v + \kappa \theta + 2 r^2 - v^2 \right) + 12v^2y_j^2 }{vy_j}{\it \delta_x^2}} u_{i,j} \\
-\frac{1}{12}{\frac { 2\,{h}^{2}{\kappa}^{2}{v}^{2}{y_j}^{2}-4\,{h}^{2}{\kappa}
^{2}\theta\,vy_j-{h}^{2}\kappa\,{v}^{3}y_j+2\,{h}^{2}{\kappa}^{2}{\theta}^
{2}-{h}^{2}\kappa\,\theta\,{v}^{2}-{h}^{2}{v}^{4}+6\,{v}^{4}{y_j}^{2}}{v^3 y_j}} {\it \delta^2_y} u_{i,j} \\
-\frac{1}{12}\,vy_j{h}^{2} \left( 2\,{\rho}^{2
}+1 \right) {\it \delta^2_x \delta^2_y} u_{i,j}
-\frac{1}{6}\,{\frac { \left( -2\,v\lambda\,\rho\,
\xi_{{B}}-y_j\rho\,{v}^{2}-\kappa\,vy_j+2\,vr\rho+\kappa\,\theta \right) {
h}^{2} }{v}{\it \delta^2_x \delta_y} } u_{i,j} \\
-\frac{1}{12}\,{\frac { \left( -4\,\kappa\,\rho\,vy_j+
4\,\rho\,\kappa\,\theta-2\,\lambda\,v\xi_{{B}}-y_j{v}^{2}+2\,rv \right) 
{h}^{2}}{v}{\it \delta_x \delta^2_y}} u_{i,j} \\
+\frac{1}{6}{\frac {   h^2\kappa \left( \rho v^2 y_j + v^2 y_j^2 - 2 r v y_j - \theta v y_j + 2 v y_j \xi_B +2 r \theta - 2\theta \xi_B \right)
-h^2 v^2 \left(\rho v- r+ \xi_B \right)  }{v^2 y_j}{\it \delta_x \delta_y}} u_{i,j} \\
+\frac{1}{12} {\frac { -{h}^{2}\kappa\,y_j v+{h}^{2}\kappa\,\theta-{h}^{2}{v}^{2}+12\,vy_j
\lambda\,\xi_{{B}}+6\,{y_j}^{2}{v}^{2}-12\,vy_j r  }{vy_j} {\it \delta_x}} u_{i,j} \\
+\frac{1}{6}\,{\frac {\kappa\, \left( -{h}^{2}\kappa\,y_j v+{h}^{2}\kappa\,
\theta-{h}^{2}{v}^{2}+6\,{y_j}^{2}{v}^{2}-6\,\theta\,vy_j \right) }{v^2 y_j} {\it 
\delta_y}} u_{i,j} \\
= f_{i,j} +\frac{h^2}{6} \frac{\rho}{v} \delta_x \delta_y f_{i,j} - \frac{h^2}{6}\ \frac{\left( v^2 +\kappa \left(v y_j -\theta \right) \right)}{v^2 y_j} \delta_y f_{i,j} -\frac{h^2}{12} \frac{ \left( 2\lambda \xi_B + 2 \rho v +v y_j - 2r \right) }{v y_j} \delta_x f_{i,j} \\+\frac{h^2}{12} \delta_x^2 f_{i,j}   +\frac{h^2}{12} \delta_y^2 f_{i,j}  .
\end{multline}
The fourth-order compact scheme \eqref{eq:HOCelliptic} considered at mesh point $(i,j)$ involves the nearest eight neighbouring meshpoints. Associated to the shape of the computational stencil, we introduce indexes for each node from zero to eight,
\[
 \begin{pmatrix}
    u_{i-1,j+1} = u_6 & u_{i,j+1} = u_2 & u_{i+1,j+1} = u_5 \\
    u_{i-1,j} = u_3 & u_{i,j} = u_0 & u_{i+1,j} = u_1 \\
    u_{i-1,j-1} = u_7 & u_{i,j-1} = u_4 & u_{i+1,j-1} = u_8
\end{pmatrix}.
\]
With this indexing the scheme \eqref{eq:HOCelliptic} is defined by 
\[\sum_{l=0}^8 \alpha_l u_l = \sum_{l=0}^8 \gamma_l f_l , \]
with the coefficients $\alpha_l $ and $\gamma_l $ given by
 \begin{align*}
\alpha_0 =&\left(  {\frac {4 {\kappa}^{2}+{v}^{2}}{12v}}- {\frac {v
 \left( 2 {\rho}^{2}-5 \right) }{3{h}^{2}}} \right) y- {\frac {2 
{\kappa}^{2}\theta+\kappa {v}^{2}+r{v}^{2}-{v}^{2}\xi_{{B}}}{3{v}^{2}}
} \\ &+ {\frac {-r\rho {v}^{3}+\rho {v}^{3}\xi_{{B}}+{\kappa}^{2}{
\theta}^{2}+{r}^{2}{v}^{2}-2 r{v}^{2}\xi_{{B}}-{v}^{4}+{v}^{2}{\xi_{{
B}}}^{2}}{3{v}^{3}y}} ,\\
\alpha_{1,3} =& \left( -\frac{v}{24}+{\frac {2\kappa\rho \pm v}{6h}}+{\frac {v
 \left( \rho-1 \right)  \left( \rho+1 \right) }{3{h}^{2}}} \right) y
               \mp
               \frac{1}{24}\kappa h+\frac{\kappa}{12}+\frac{r}{6}-\frac{\xi_{{B}}}{6}
               \pm {\frac {\kappa \rho \theta-rv+v\xi_{{B}}}{3vh}}
   \\ 
&+ {\frac {1}{y} \left( \pm {\frac { \left( 
 \kappa \theta-{v}^{2} \right) h}{24v}}-{\frac {-2 r\rho v+2 
\rho v\xi_{{B}}+\kappa \theta+2 {r}^{2}-4 r\xi_{{B}}-{v}^{2}+2 {
\xi_{{B}}}^{2}}{12v}} \right) }, \\
\alpha_{2,4} =& \left( {\frac {{\kappa}^{2}}{6v}}+{\frac { \mp \rho v \pm 2\kappa
}{6h}}+{\frac {v \left( \rho-1 \right)  \left( \rho+1 \right) }{3{h
}^{2}}} \right) y \mp {\frac {{\kappa}^{2}h}{12v}}+{\frac {
\kappa  \left( 4 \kappa \theta+{v}^{2} \right) }{12{v}^{2}}}-{
\frac {-r\rho v + \rho v\xi_{{B}}+\kappa \theta}{3vh}} \\ 
&+{\frac {1}{y}
 \left( {\frac {\kappa  \left( \kappa \theta-{v}^{2} \right) h
}{12{v}^{2}}}-{\frac { \left( 2 \kappa \theta+{v}^{2} \right) 
 \left( \kappa \theta-{v}^{2} \right) }{12{v}^{3}}} \right) } ,\\
\alpha_{5,7} =& \left( -\frac{\kappa}{24} \pm {\frac { \left( 2 \rho+1 \right)  \left( 2 
\kappa+v \right) }{24h}}-{\frac {v \left( \rho+1 \right)  \left( 2
 \rho+1 \right) }{12{h}^{2}}} \right) y+{\frac {\kappa  \left( 
\rho v+2 r+\theta-2 \xi_{{B}} \right) }{24v}} \\
&\mp {\frac { \left( 2
 \rho+1 \right)  \left( \kappa \theta+rv-v\xi_{{B}} \right) }{12vh}}
  - {\frac {-\rho {v}^{3}+2 \kappa r\theta-2 \kappa \theta \xi_
{{B}}-r{v}^{2}+{v}^{2}\xi_{{B}}}{24{v}^{2}y}}, \\
\alpha_{6,8} =& \left( \frac{\kappa}{24} \mp {\frac { \left( 2 \rho-1 \right)  \left( 2 
\kappa-v \right) }{24h}}-{\frac {v \left( 2 \rho-1 \right) 
 \left( \rho-1 \right) }{12{h}^{2}}} \right) y-{\frac {\kappa 
 \left( \rho v+2 r+\theta-2 \xi_{{B}} \right) }{24v}}\\
& \pm {\frac {
 \left( 2 \rho-1 \right)  \left( \kappa \theta-rv+v\xi_{{B}}
 \right) }{12vh}} +{\frac {-\rho {v}^{3}+2 \kappa r\theta-2 
\kappa \theta \xi_{{B}}-r{v}^{2}+{v}^{2}\xi_{{B}}}{24{v}^{2}y}} ,
 \end{align*}
 and
\begin{align*}
\gamma_0 &= 2/3, \quad \gamma_{1,3} = \frac{1}{12} \mp \frac{h}{24} \pm {\frac { \left( -\rho v+r-\xi_{{B}} \right) h}{12vy}},\quad
\gamma_{2,4} = \frac{1}{12} \mp {\frac {\kappa h}{12v}} \pm {\frac { \left( \kappa \theta-{v}^{2} \right) h}{12{v}^{2}y}}, \\
\gamma_5 &= \gamma_7 =\frac{\rho}{24}, \quad \gamma_6 = \gamma_8 = -\frac{\rho}{24} .
\end{align*}
When multiple indexes are used with $\pm$ and $\mp$ signs, the first index corresponds to the upper sign.
\subsubsection{Extension to the parabolic problem}

\noindent To extend the above approach to the parabolic problem we
replace $f(x,y) $ in \eqref{eq:Lelliptic} by the time derivative. We
consider the class of two time step methods. By differencing at $
t_{\mu} = ( 1 - \mu ) t_n + \mu t_{n+1} $, where $ 0 \leq \mu \leq 1 $
and the superscript $n$ denotes the time level, we yield a set of
integrators including the forward and backward Euler scheme, for $ \mu=0  $ and $  \mu=1 $, respectively, and the Crank-Nicolson scheme $(\mu=1/2)$. By defining $ \delta_t^+ u^n = \frac{u^{n+1}-u^{n}}{k} $, the resulting fully discrete difference scheme of node $(i,j)$ at the time level $n$ becomes
\[\sum_{l=0}^8 \mu \alpha_l u^{n+1}_l + (1-\mu) \alpha u_l^n = \sum_{l=0}^8 \gamma_l \delta^+_t u_l^n ,\]
which can be written as
\[\sum_{l=0}^8 \beta_l u_l^{n+1} = \sum_{l=0}^8 \zeta_l  u_l^n .\]
with the coefficients $\beta_l $ and $\zeta_l $ given by
\begin{align*} \beta_{0} =& (((2y_j^2 - 8)v^4 +((-8\kappa - 8r +
                            8\xi_B )y_j - (8r + 8 \xi_B ) \rho ) v^3 +
                            ( 8\kappa^2 y_j^2 + 8 r^2 -16 r \xi_B + 8
                            \xi_B) v^2 \\ &-16 \kappa^2 \theta v y_j +8
  \kappa^2 \theta^2 ) \mu k +16v^3y_j)h^2 + (16\rho^2 + 40 )y_j^2 v^4
  \mu k ,\\
\beta_{1,3} =& \pm (( \kappa \theta v^2 - v^4 - \kappa y_j v^3 )\mu k
               - (y_j +2\rho) v^3 + 2v^2r  - 2 v^2 \xi_B ) h^3 + (((
               -y_j^2 +2 )v^4 \\ &+((4r - 4\xi_B +2\kappa ) y_j + 4\rho
  r - 4\rho \xi_B  )v^3 - (2\kappa \theta + 4 r^2 - 4 xi_B^2 + 8 r
  \xi_B  )v^2 )\mu k + 2v^3 y_j ) h^2  \\ &\pm (4v^4y_j^2 + (-8y_j^2 \kappa
  \rho - 8 y_j r + 8 y_j \xi_B  ) v^3+ 8y_j \kappa \theta \rho
  v^2 ) \mu k h  + ( 8 \rho^2 - 8) y_j^2 v^4 \mu k,\\
\beta_{2,4} =& \pm (( 2 \kappa^2 \theta v - 2 \kappa^2 v^2 y_j - 2v^3
               \kappa ) \mu k - 2v^2 y_j \kappa + 2v \kappa \theta  -
               2v^3)h^3 + ((2v^4 + 2 \kappa y_j v^3 \\ &+( -4\kappa^2
               y_j^2 + 2 \kappa \theta ) v^2 + 8 \kappa^2 v y_j - 4
  \kappa^2 \theta^2 ) \mu k + 2v^3 y_j )h^2 \pm (( 8y_j^2 \kappa +
  8y_j\rho r - 8y_j \rho \xi_B )v^3 \\& -4v^4 y_j^2 \rho - 8v^2 y_j \kappa
  \theta ) \mu k h  + (8 \rho^2 - 8 )y_j^2 v^4 \mu k, \\
 \beta_{5,7} =& (( v^4 \rho + (-y^2 \kappa + \kappa y_j \rho +r -
                \xi_B ) v^3 +(\theta +2r - 2\xi_B ) \kappa y_j v^2 -
                2r \kappa \theta v + 2\xi_B \kappa \theta v ) \mu k +
                v^3 \rho y_j) h^2 \\  &\pm (( 2\rho +1 ) y_j^2 v^4 + ((
  2 +4\rho )\kappa y_j^2 +(-2r + 2\xi_B - 4\rho r + 4\rho \xi_B)
  y_j)v^3 + (-4\theta \rho - 2\theta )\kappa y_j v^2 ) \mu k h \\ & +
  (-4\rho^2 -6\rho - 2) y_j^2 v^4 \mu k , \\
\beta_{6,8} =& (( -v^4 \rho + (y^2 \kappa - \kappa y_j \rho -r + \xi_B
               ) v^3 +(-\theta -2r + 2\xi_B ) \kappa y_j v^2 + 2r
               \kappa \theta v - 2\xi_B \kappa \theta v ) \mu k - v^3
               \rho y_j) h^2 \\ & \pm (( 2\rho -1 ) y_j^2 v^4 + (( 2 -
  4\rho )\kappa y_j^2 +(2r - 2\xi_B - 4\rho r + 4\rho \xi_B) y_j)v^3 +
  (4\theta \rho - 2\theta )\kappa y_j v^2 ) \mu k h \\  &+ (-4\rho^2
  +6\rho - 2) y_j^2 v^4 \mu k , 
\end{align*}  
and
\begin{align*} \zeta_0 =& 16v^3y_j h^2 + (1-\mu) k ((( 8-2y_j^2 )v^4 +
  ((8\kappa +8r - 8 \xi_B )y_j +8\rho r - 8 \rho \xi_B)v^3 \\& + (-8r^2
  - 8\xi_B^2 +16r\xi_B -8\kappa^2 y_j^2 )v^2 +16 \kappa^2 \theta v y_j
  - 8 \kappa^2 \theta^2 )h^2 +(-40+16\rho^2)y_j^2v^4), \\
\zeta_{1,3} = &\pm(2r - 2 \xi_B - (y_j +2\rho)v)v^2h^3 + 2v^3 y_j h^2
                +(1-\mu) k(\pm (v\kappa y_j +v^2 - \kappa \theta )v^2
                h^3  \\& +( v^2 y_j^2 - (4r + 4 \xi_B + 2 \kappa ) v
                         y_j +4r^2  + 4 \xi_B^2  + 2 \kappa \theta + 2
                         v y_j - 4\rho v r + 4 \rho v \xi_B ) v^2 h^2
  \\&  \pm ((-4v +8 \kappa \rho )v^3 y_j^2 + (-8 \kappa \theta \rho
      +8 v r - 8 v \xi_B ) v^2 y_j ) h   + ( 8 v^2 - 8 v^2 \rho^2 )
      v^2 y_j^2 ) , \\
\zeta_{2,4} = &\pm( 2 v \kappa \theta - 2v^2 y_j \kappa - 2v^3 ) h^3 +
                2v^3 y_j h^2 + (1-\mu) k(\pm (v\kappa y_j +v^2 -
                \kappa \theta ) v^2 h^3 \\ &+ (v^2 y_j^2 -(4r+2\kappa)vy_j
  + 2 \kappa \theta ( 2\kappa \theta - v^2) - 2v^4)h^2 \pm ((-8
       v^3 \kappa + 4 v^4 \rho )y_j^2  \\ &+(8 \kappa \theta v^2 -8 v^3
       \rho r )y_j) h  +(-8v^4 \rho^2 +8v^4)y_j^2 ),\\
 \zeta_{5,7} =& v^3 \rho y_j h^2  + (1 - \mu) k((v^3y_j^2 \kappa -
               v(v\kappa \theta + 2 r \kappa v - 2 \xi_B \kappa v +
               \kappa v^2 \rho ) y_j ) \\ &- v(v^2 r - 2v^2 \xi_B - 2r
               \kappa \theta + 2 \xi_B \kappa \theta +v^3  \rho )) h^2
  \pm (-v(2v^3 \rho + v^3 + 4 \kappa v^2 \rho +2 v^2 \kappa
       )y_j^2 \\ &+ v (2 v \kappa \theta + 4 v \kappa \theta \rho + 4v^2
       \rho r + 4v^2 \rho \xi_B + 2v^2 r + 2v^2 \xi_B) y_j ) h +v(2v^3
       + 6v^3 \rho +4v^3 \rho^2)y_j^2), \\ 
\zeta_{6,8} = & -v^3 \rho y_j h^2 + (1-\mu ) k (( -v^3 y_j^2 \kappa +
                v (v \kappa \theta + 2r \kappa v - 2 \xi_B \kappa v +
                \kappa v^2 \rho) y_j \\ &+v ( v^2 r - v^2 \xi_B - 2 r
                \kappa \theta + 2 \xi_B \kappa \theta  + v^3 \rho )) \pm (v(-2v^3 \rho +v^3 + 4 \kappa v^2 \rho - 2v^2 \kappa )y_j^2\\ 
& +v(2v\kappa \theta - 4v \kappa \theta \rho + 4v^2 \rho r - 4 v^2
    \rho \xi_B - 2v^2 r - 2 v^2 \xi_B ) y_j) h +v(2v^3 -6v^3 \rho + 4v^3 \rho^2) y_j^2).
\end{align*}
Where multiple indexes are used with $\pm$ and $\mp$ signs, the first
index corresponds to the upper sign. The Crank-Nicholson scheme is
used by setting $ \mu = 1/2 $, yielding a 
scheme which is second-order accurate in time and fourth-order accurate in space.

\subsection{Integral operator}

\noindent
After the initial transformation of variables we have the integral operator in the following form,
\[ L_I = \lambda \int^{+\infty}_{-\infty} \tilde u(x + z , y , \tau ) \tilde p(z) \,\mathrm{d}z , \]
where the probability density function, $\tilde p(z)$ is given by 
 \[ \tilde p(z)=\frac{1}{\sqrt{2\pi} z v}e^{-\frac{(\text{log}(z)-\gamma)^2}{2v^2}}.\]
We make a final change of variables $ \zeta = x+z $ with the intention of studying the value of the integral at the point  $ x_i $, 
\begin{multline} 
\label{eq:int}
I_i =  \int^{+\infty}_{-\infty} \tilde u( \zeta , y , \tau ) \tilde
p(\zeta - x_i) \,\mathrm{d}\zeta = \int^{x_{max}}_{x_{min}} \tilde u( \zeta , y ,
\tau ) \tilde p(\zeta - x_i) \,\mathrm{d}\zeta + \int^{\infty}_{x_{max}} \tilde
u( \zeta , y , \tau ) \tilde p(\zeta - x_i) \,\mathrm{d}\zeta \\ +
\int^{x_{min}}_{-\infty} \tilde u( \zeta , y , \tau ) \tilde p(\zeta -
x_i) \,\mathrm{d}\zeta .
\end{multline}

\subsubsection{Simpson's rule}

\noindent To estimate the integral we require a numerical integration method of
high order to match our finite difference scheme, we choose to use the composite Simpson's rule, defined as
\[ \int_a^b f(x) \,\mathrm{d} x \approx \frac{h}{3} \left[f(x_0) + 2 \sum_{j=1}^{n/2-1} f(x_{2j}) + 4 \sum_{j=1}^{n/2} f(x_{2j-1}) +f(x_n) \right]  .\]
The error committed by the composite Simpson's rule is bounded by
\[ \frac{h^4}{180}(b-a) \max_{\xi \in [a,b]} |f^{4}(\xi)| .\]
Through the choice of the interval $(x_{min} , x_{max} ) $ we can assure that the integrals outside this range are of negligible value. Allowing the integral to be evaluated using Simpsons rule on a equidistant grid in $x$ with spacing $ \Delta x $ and $m_x $ grid-points in $(x_{min} , x_{max} ) $, where each interval has length mesh-size $h/2$. Equation \eqref{eq:int} can now be written as,
\begin{multline*} 
I_i  \approx \int^{x_{max}}_{x_{min}} \tilde u( \zeta , y , \tau ) \tilde p(\zeta - x_i) \, \mathrm{d}\zeta \\ \approx \frac{\Delta x}{3}  \sum_{j=1}^{\frac{m_x}{2}} \bigl[ \tilde u( \zeta_{2j-2} , y , \tau ) \tilde p(\zeta_{2j-2}- x_i) + 4\tilde u( \zeta_{2j-1} , y , \tau ) \tilde p(\zeta_{2j-1}- x_i) \\+\tilde u( \zeta_{2j} , y , \tau ) \tilde p(\zeta_{2j}- x_i)   \bigr] = \tilde I_i.
\end{multline*}
This computation is calculated explicitly at each time-step by the matrix-vector equation, $ \tilde I = W_{m_x} \tilde u $, defined as follows,
\begin{align*} \tilde I = \left( \tilde I_1 \quad \tilde I_3 \quad
  ... \quad \tilde I_{{m_{x-1}}/2} \quad \tilde I_{{m_x}/2} \right)
  ^\top ,  \quad \tilde u = \left( \tilde u_1 \quad \tilde u_3 \quad
  ... \quad \tilde u_{{m_{x-1}}/2} \quad \tilde u_{{m_x}/2}
  \right)^\top ,
\end{align*}
\[W_{m_x}
 =
 \begin{bmatrix}
    \tilde p(\zeta_{0}- x_0)  & 4\tilde p(\zeta_{1}- x_0) & 2\tilde p(\zeta_{2}- x_0) & \dots & \tilde p(\zeta_{mx}- x_0) \\
 \tilde p(\zeta_{0}- x_1)  & 4\tilde p(\zeta_{1}- x_1) & 2\tilde p(\zeta_{2}- x_1) & \dots & \tilde p(\zeta_{mx}- x_1) \\
\vdots & \vdots & \vdots & \ddots & \vdots \\
\tilde p(\zeta_{0}- x_{mx})  & 4\tilde p(\zeta_{1}- x_{mx}) & 2\tilde p(\zeta_{2}- x_{mx}) & \dots & \tilde p(\zeta_{mx}- x_{mx}) 
\end{bmatrix}.
\]

The integral operator $L_I$ is estimated over $( x_{min}, x_{max} )$
using Simpson's rule. The tails could be discarded as they are assumed
to be of negligible value for sufficiently small (large) choice of
$x_{min}$ ($x_{max}$). A direct result of this approach would be the
necessity to compute the option price over a wider domain than
practically relevant.
To alleviate this issue we assume that the option price follows the
payoff function outside of the range $( x_{min}, x_{max} )$, and
approximate the tails by
the following integrals 
\[ \int^{\infty}_{x_{max}} \tilde u( \zeta , y , \tau ) \tilde p(\zeta) \,\mathrm{d}\zeta \approx \int^{\infty}_{x_{max}} \text{max}(1-\text{exp}(\zeta),0) \tilde p(\zeta) \,\mathrm{d}\zeta , \]

\[ \int^{x_{min}}_{-\infty} \tilde u( \zeta , y , \tau ) \tilde p(\zeta) \,\mathrm{d}\zeta \approx \int^{x_{min}}_{-\infty}  \text{max}(1-\text{exp}(\zeta),0) \tilde p(\zeta) \,\mathrm{d}\zeta . \]

The value of the first of these integrals is trivial as the payoff
function for the Put option is zero in the region $(x_{max}, +\infty) $. We estimate the
second integral using Simpson's rule on an equal-sized adjacent
equidistant grid to our original grid.

\subsection{Time discretisation for IMEX method}

\noindent Having set the framework for the discretisation of the operators $L_D$
and $L_I$, we now introduce the implicit-explicit method,
\[ \sum_{l=0}^8 \beta_l u^{n+1} = \sum_{l=0}^8 \zeta_l  \left( 1 + \frac{3 \Delta \tau}{2} L_I \right) u^n - \sum_{l=0}^8 \zeta_l  \left( \frac{\Delta \tau}{2} L_I \right) u^{n-1}. \]

\section{Initial condition and boundary conditions}
\label{sec:initboundcond}

\subsection{Initial condition}

\noindent The initial condition is given by the transformed payoff
function of the Put option,
\[ u(x,\sigma ,0)= \max (1- \exp(x),0), \quad x \in \mathbb{R}, \;\sigma > 0 .\]
To maintain the order of the scheme we smooth this function around zero, this follows from \cite{Kreiss} which states that we
cannot expect to achieve fourth order convergence if the initial
condition is not sufficiently smooth. In \cite{Kreiss}  suitable
smoothing operators are defined in the Fourier space. Since the order
of convergence of our high-order compact scheme is four we follow
\cite{DuHe15} and select the smoothing operator $ \phi_4 $, given by its Fourier transform
\[ \phi_4 (\omega) = \left( \frac{\sin\left(\omega/2 \right) }{ \omega/2 } \right)^4 \left[1 +\frac{2}{3} \sin^2\left( w/2 \right) \right] . \]
This leads to the smoothed initial condition
\[ \tilde{u}_0 (x_1, x_2) = \frac{1}{h^2} \int_{-3h}^{3h}  \int_{-3h}^{3h} \phi_4 \left( \frac{x}{h} \right)  \phi_4 \left( \frac{y}{h} \right) u_0 (x_1 - x , x_2 - y) \, \mathrm{d} x  \mathrm{d} y .\] 
As $h\to 0$, this smoothed initial condition converges to the original
initial condition. The results in \cite{Kreiss} prove high-order
convergence of the approximation to the smoothed problem to the true
solution of \eqref{eq:Ptransf}.

Note that in \cite{DuFo12} a Rannacher style
smoothing start-up \cite{Rannacher} is used with four fully implicit
quarter time steps. In our
experiments with the high-order compact scheme we notice no benefit by employing such a start-up, and use
the Crank-Nicolson time stepping throughout. Since the coefficients in
\eqref{eq:Ptransf} do not depend on time, we are required to build up the
discretisation matrices for the new scheme only once. They can then be
$LU$-factorised once, and the factorisation can be used in each time
step, leading to a highly efficient scheme.

\subsection{Boundary conditions}

\noindent We impose artificial boundary conditions as follows. Due to the compactness of the scheme, the Dirichlet boundary conditions are considered without introduction of numerical error by imposing
\[ u^n_{-N,j} = 1- e^{r t_n - N h}  ,  \quad u_{+N,j}^n = 0, \quad j=0, ... , M. \]
At the other boundaries we impose homogeneous Neumann boundary
conditions, these require more attention as no value is prescribed,
therefore, they must be set by extrapolation from values in the
interior. Here the introduction of numerical error must be negated by
choice of an extrapolation formulae of order high enough not to affect
the overall order of accuracy. We choose the following extrapolation
formulae:
\begin{align*} 
u_{i,0}^n &= 4 u_{i,1}^n - 6 u_{i,2}^n + 4 u_{i,3}^n - u_{i,4}^n +\mathcal{O}(h^4),
  \quad i=-N +1 , ... , N-1, \\
 u_{i,M}^n &= 4 u_{i,M-1}^n - 6 u_{i,M-2}^n + 4u_{i,M-3}^n -
  u_{i,M-4}^n +\mathcal{O}(h^4), \quad i=-N +1 , ... , N-1 .
\end{align*}

\section{A finite element method for comparison}
\label{sec:FEM}



\noindent
In addition to standard, second-order finite difference methods we will
compare our new scheme to different finite element methods. In this
short section we briefly state the variational formulation of the PIDE
problem.

We can rewrite the equation for the differential operator $L_D$ in divergence form,
\begin{equation*}
u_{\tau} - \text{div} \left( A \nabla u \right) + b \cdot \nabla u = 0 ,
\end{equation*}
where the coefficients $A$ and $b$ are given by
\begin{equation*}
A = \frac{1}{2}vy 
\begin{bmatrix}
    1 & \rho \\
    \rho & 1 
\end{bmatrix}
\ ,
\quad b =
\begin{bmatrix}
   \frac{1}{2} vy -  r + \lambda \xi_B - \frac{v \rho}{2} \\
    -\kappa \frac{\left( \theta - vy \right)}{v} - \frac{v}{2}
\end{bmatrix}.
\end{equation*}

To solve this problem using finite elements we produce a variational
formulation, which requires multiplying by suitable test functions
$\phi$ and integrating over the domain $\Omega$.



Mirroring the approach defined in Section~\ref{sec:impexp}, we employ
an IMEX discretisation with the integral operator, $L_I$, being
computed using the Simpson's rule. 
We have the following Crank-Nicholson scheme,
\begin{multline*}
\left( \int_{\Omega} u^{n+1} \phi \ \text{d}x \text{d}y + \left[
    \int_{\Omega} A \nabla u^{n+1} \cdot \nabla \phi \ \text{d}x
    \text{d}y   +  \int_{\Omega} b \cdot  \nabla u^{n+1} \phi \
    \text{d}x \text{d}y  \right] \frac{\Delta \tau}2 \right) = \\
\left( \int_{\Omega} u^{n} \phi \ \text{d}x \text{d}y + \left[
    \frac12\int_{\Omega} A \nabla u^{n} \cdot \nabla \phi \ \text{d}x
    \text{d}y   + \frac12 \int_{\Omega} b \cdot  \nabla u^{n} \phi \
    \text{d}x \text{d}y + \frac{3}{2} L_I u^n - \frac{1}{2} L_I
    u^{n-1} \right] \Delta \tau \right).
\end{multline*}

\section{Numerical experiments}
\label{sec:num}

\noindent 
In our numerical experiments we compare the performance of two finite
difference schemes, a standard, second-order central difference
scheme and the new HOC scheme, against two variants of the finite
element approach presented in the previous section, using Lagrange
elements with linear ($p=1$) and 
quadratic ($p=2$) 
polynomial basis
functions on quadrilateral meshes.
While a finite element method with cubic basis functions ($p=3$) would
be expected to give a similar numerical convergence order as the high-order compact
scheme, the number of degrees of freedom would increase substantially,
and make this approach less viable, see also comments below in
Section~\ref{sec:numconv}.

Both finite difference schemes are implemented in C++. 
For our numerical experiments with finite elements we use the FEniCS
FEM solver. FEniCS is a popular open-source platform which allows
users quickly to obtain efficient FEM code for solving partial
differential equations. The code is written in Python 3.5 and utilises the inbuilt
packages of NumPy and SciPy to improve efficiency.

We measure the convergence,
computational time, number of unknowns and the memory usage for each method. As a
separate study we compare the stability of the new HOC finite
difference scheme against a standard, second-order central difference
scheme.

\begin{figure}[H]
\epsfig{file=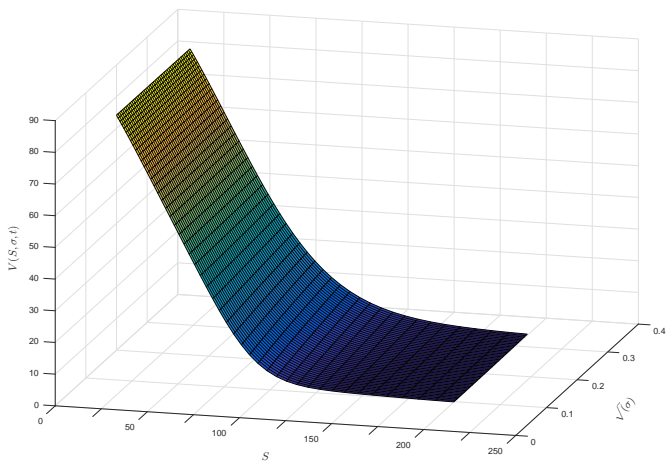, width=10cm}
\caption{ Price of a European Put Option.} 
\label{f:price}
\centering
\end{figure}

Below we present Figure~\ref{f:price} which shows the price of a European put
option plotted against the volatility $\sqrt{\sigma}$ and the asset
price $S$. The default parameters used for the numerical experiments are given in Table~\ref{t:params}.

\begin{table}[h]
\centering
\begin{tabular} { c l }
\toprule 
Parameter & Value \\
\midrule
Strike Price & $K=100$ \\
Time to maturity & $T=0.5$ \\
Interest rate & $r=0.05$ \\
Volatility of volatility & $ v=0.1 $ \\
Mean reversion speed & $ \kappa = 2 $ \\
Long-run mean of $\sigma$ & $ \theta =0.01 $ \\
Correlation & $\rho=-0.5$ \\
Jump Intensity & $ \lambda = 0.2 $ \\
\bottomrule
\end{tabular} \\
\vspace{2mm}
\captionof{table}{Default parameters for numerical simulations.}
\label{t:params}
\end{table}

\subsection{Numerical convergence}
\label{sec:numconv}

\noindent We perform a numerical study to evaluate the rate of convergence of the schemes. We refer to both the $ l_2 $-error $\epsilon_2$ and the $ l_{\infty} $-error $\epsilon_\infty $ with respect to a numerical reference solution on a fine grid with $h_{\text{ref}} =0.025. $ With the parabolic mesh ratio $k/h^2$ fixed to a constant value we expect these errors to converge as $ \epsilon= Ch^m $ for some $m$ and $C$ which represent constants. From this we generate a double-logarithmic plot $\epsilon$ against $h$ which should be asymptotic to a straight line with slope $m$, thereby giving a method for experimentally determining the order of the scheme. 

We compare the new HOC scheme to the finite element approach from
Section~\ref{sec:FEM}
 (with polynomial orders $p=1,2$) 
and a
standard, second-order central finite difference scheme. The
second-order finite difference scheme requires a Rannacher style start-up
\cite{Rannacher} which involves starting by four quarter fully
implicit Euler steps to combat stability issues \cite{Giles}.  

\begin{figure}[H]
	\centering
	\epsfig{file=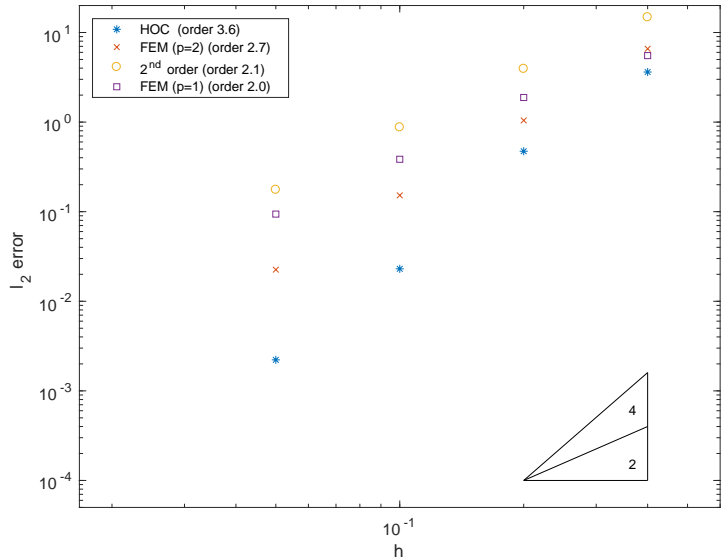, width=11cm}
	\caption{ $l_2$-error in option price taken at mesh-sizes
          $h=0.4,0.2,0.1,0.05.$}
\label{f:l2vsh}
\end{figure}

\begin{figure}[H]
	\centering
	\epsfig{file=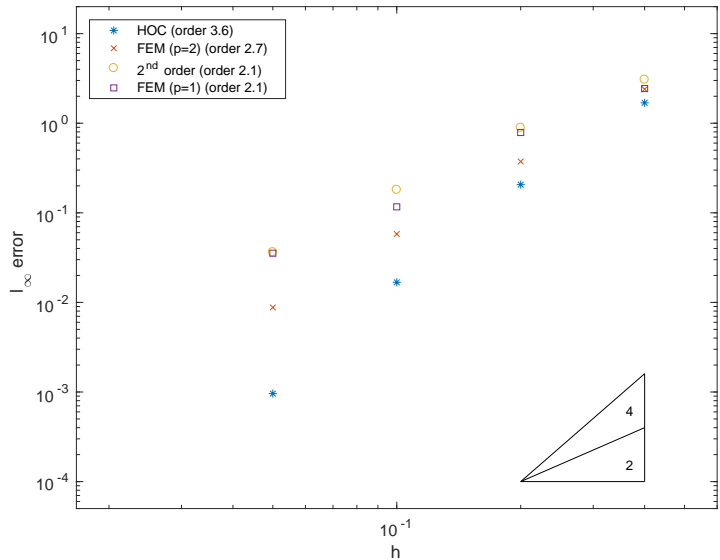, width=11cm}
	\caption{ $l_\infty$-error in option price taken at mesh-sizes
          $h=0.4,0.2,0.1,0.05.$}
\label{f:linfvsh}
\end{figure}

These numerical convergence results are
included in Figure~\ref{f:l2vsh} for the $ l_2 $-error $\epsilon_2$ and
Figure~\ref{f:linfvsh} for the $ l_{\infty}$-error $\epsilon_\infty $. 
The numerical convergence orders are estimated from the slope of a least squares
fitted line. 

We observe that the numerical convergence orders are consistent with the
theoretical order of the schemes. We note that the finite element
approach with 
$p=2$ achieves a rate close to three 
whereas the new high-order compact
scheme has convergence rates close to four. 
With a finite element method with cubic basis functions ($p=3$) one
would be able to match the fourth order of the high-order compact scheme, but only at the expense of solving a
much larger system, due to the much larger number of degrees of
freedom for $p=3$. For example, on a mesh with $h=0.05$ the cubic
finite element method would employ 58081 degrees of freedom, almost
ten times more than the high-order compact scheme on the same mesh. 


\subsection{Computational efficiency comparison}

\noindent We conduct an efficiency comparison between
the new high-order scheme, a standard second-order discretisation and
the finite element method with polynomial basis order 
$p=1$ and 
$p=2$. The finite element
methods employ quadrilateral meshes to allow for better comparison
with the finite difference methods.

We compare the computational time to obtain a given accuracy, taking
into account matrix setups, factorisation and boundary condition
evaluation. The timings depend obviously on technical details of the
computer as well as on specifics of the implementation. Care was taken
to implement both finite difference schemes in an efficient and consistent manner, using standard libraries where possible, to
avoid unnecessary bias in the results.
Direct comparison of computational times with the Python based FEM
schemes are difficult, but still give an indication what can be
achieved with a standard `black-box' solver. 
All results were computed on the same laptop computer (2015 MacBook Air 11''). 

Since the coefficients in \eqref{eq:Ptransf} do not depend on time, we
are required to build up the discretisation matrices for the new
scheme only once (twice for the second-order scheme with Rannacher
start-up). 
The new scheme requires only one initial $LU$-factorisation of a
sparse matrix. This factorisation is then
employed in each time step, leading to a highly efficient scheme.
Further efficiency gains are obtainable by parallelisation or GPU computing.

The results are shown below in Figure~\ref{f:eff}. The mesh-sizes
used for this comparison are $h=0.4$, $h=0.2$, $h=0.1$ and $h=0.05$,
with the reference mesh-size used being $h_{\mathrm{ref}}=0.025$. From this
comparison it is clear to note that the high-order compact scheme
achieves higher accuracy with less computational time at all
mesh-sizes. The improvement in computational time over the
second-order finite difference scheme can be partly
attributed to the absence of the Rannacher start-up which
requires the additional construction and factorisation of a sparse matrix populated with
coefficients for the implicit Euler steps. 

The finite element method with $p=1$ has comparable results for both
computational time and $l_2$-error to the second-order finite difference
scheme, however, for 
$p=2$ the computational time for the finite element
method increases substantially with the size of the linear system to
be solved. 

Table~\ref{t:comp} summarises more detailed results of the numerical comparison.
The number of degrees of freedom for all schemes are shown in the third
column. The standard finite difference scheme and the linear FEM use
the same number of unknowns. It is noticeable that the HOC scheme,
unlike the high-order FEM approach with 
$p=2$, achieves high-order convergence
without requiring additional unknowns.
As a result the HOC scheme is very
parsimonious in terms of computational effort and memory requirements.

The memory requirements are an important factor in numerical
computations. Direct comparisons of memory usage between the C++
implementations of the finite difference schemes and the `black box' FEniCS FEM
approaches are not viable. Moreover, FEniCS allocates already a rather
large amount of memory at the coarsest mesh with $h=0.4$.
Hence, rather than looking at total memory used, we report the memory usage at 
each subsequent refinement as the extra memory required to the base mesh size
$h=0.4$. The results demonstrate both the improvements of the HOC scheme over 
the second-order alternative and also the greater memory required to achieve
comparable convergence with the finite element methods.

\begin{figure}
	\centering
	\epsfig{file=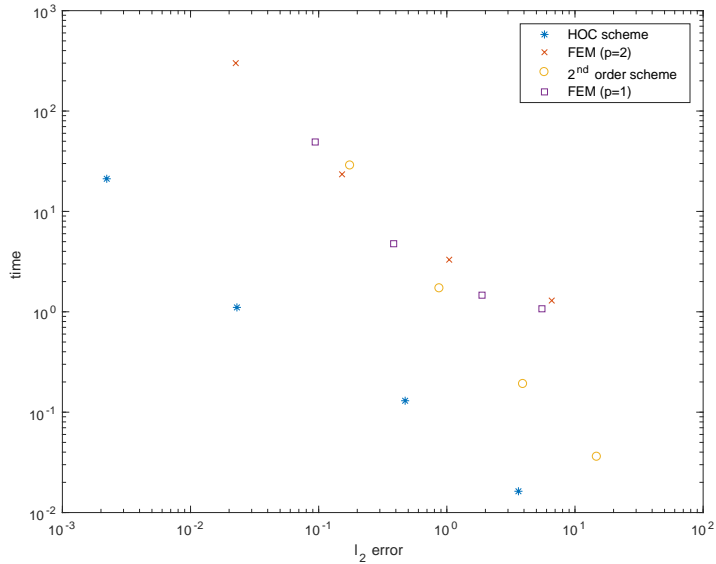, width=11cm}
	\caption{Computational speed comparison taken at mesh-sizes
          $h=0.4,0.2,0.1,0.05$.}
\label{f:eff}
\end{figure}

\begin{table}
        \centering 
        \begin{tabular}{c c c c c c c}
	\toprule
            Scheme & $h$  & DOF & $l_2$-error & $l_{\infty} $-error & time (s) & memory (kB) \\
	\midrule
	\multirow{4}{*}{HOC} & 0.4 & 121& 3.6201 & 1.6891 & 0.016 & 6916 \\ 
    	 & 0.2 & 441& 0.4728 & 0.2063 & 0.130 & +1060 \\ 
	 & 0.1 & 1681& 0.0230 & 0.0168 & 1.106 & +5536 \\ 
	 & 0.05 & 6561& 0.0022 & 0.0009 & 21.145 & +18284 \\  
	\midrule
	\multirow{4}{*}{FEM $(p=2)$} & 0.4 & 441& 6.5837 & 2.3944 & 1.294 & 123128 \\ 
    	 & 0.2 & 1681& 1.0438 & 0.3737 & 3.304 & +1780 \\ 
	 & 0.1 & 6561& 0.1522 & 0.0581 & 23.426 & +8268 \\ 
	 & 0.05 & 25921& 0.0225 & 0.0088 & 300.019 & +40828 \\  
	\midrule
	\multirow{4}{*}{FD} & 0.4 & 121& 14.8087 & 3.0653 & 0.036 & 6948 \\ 
    	 & 0.2 & 441& 3.9321 & 0.8913 & 0.191 & +1772 \\ 
	 & 0.1 & 1681& 0.8751 & 0.1806 & 1.715 & +8384 \\ 
	 & 0.05 & 6561& 0.1758 & 0.0364 & 28.706 & +23064 \\  
	\midrule
	\multirow{4}{*}{FEM $(p=1)$} & 0.4 & 121 & 5.5209 & 2.4373 & 1.072 & 123276 \\ 
    	 & 0.2 & 441 & 1.8816 & 0.7876 & 1.462 & +192 \\ 
	 & 0.1 & 1681 & 0.3846 & 0.1166 & 4.727 & +2052 \\ 
	 & 0.05 & 6561 & 0.0940 & 0.0354 & 49.171 & +8176 \\  
\bottomrule
        \end{tabular}
        \captionof{table}{Performance results for the HOC,
          second-order FD and FEM $(p=1,2)$ schemes. Comparison
          for computational time and memory usage between the finite
          difference schemes (HOC and second-order) and the FEM schemes
          ($p=1,2$) are only indicative since implementations are
          different.  Note that rather than total
          memory usage, increases in
memory usage at each subsequent refinement from the base mesh size
$h=0.4$ are given for each scheme.}
\label{t:comp}
\end{table}

\subsection{Numerical stability analysis}

\noindent To assess the stability of the scheme we present a numerical stability
analysis. We propose to test to what extent the parabolic mesh ratio
$k/h^2$ impacts the convergence of the scheme. If the effect is
minimal this will allow numerically regular solutions to be obtained
without restriction on the time step-size. We proceed to compute
numerical solutions for varying values of the parabolic mesh ratio $
k/h^2$ and the mesh width $h$, then plot these against the associated
$l_2$-errors to detect stability restrictions depending on
$k/h^2$. This numerical test is performed for both the high-order and
the second-order schemes, with the results shown in Figure~\ref{fig:stest1} and Figure~\ref{fig:stest2}
respectively. We use default parameters from Table~\ref{t:params}, and vary the
parabolic mesh size from $0.1$ to $1$ in increments of $0.1$. Note the
difference in the error scales between the two schemes.

For both schemes the error increases gradually as the parabolic mesh
ratio and $h$ are increased.
We note that for the second-order scheme the contour plot of the error
may indicated some mild condition on the time stepping, the effect
being stronger for larger mesh size $h$. The high-order scheme also features a mild
dependence on the parabolic mesh ratio. Although there is no apparent
stability restriction, it appears that values of the parabolic mesh
ratio below and close to $0.5$ are most useful.
We attribute this dependence of the scheme to the parabolic mesh ratio
as a consequence of the implicit-explicit nature of the scheme. For
the present option pricing problem, the restriction on the time
stepping for the new scheme is not severe, since the discretisation matrices do not
change in time (the coefficients in the partial integro-differential
equation \eqref{eq:Ptransf} do not depend on time). Hence, the sparse matrix
factorisation is performed only once, and additional time steps do
not require additional factorisations to solve the problem.

\begin{minipage} {0.45\textwidth}
\begin{figure}[H]
	\centering
	\epsfig{file=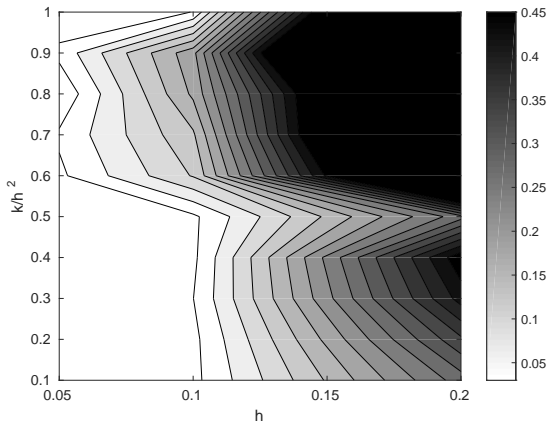, width=8cm}
	\captionof{figure}{Contour plot of the $l_2$-error for
          the HOC scheme.} 
	\label{fig:stest1}
	\end{figure}
\end{minipage}%
\hspace{0.05\textwidth}
\begin{minipage} {0.45\textwidth}
\begin{figure}[H]
	\centering
	\epsfig{file=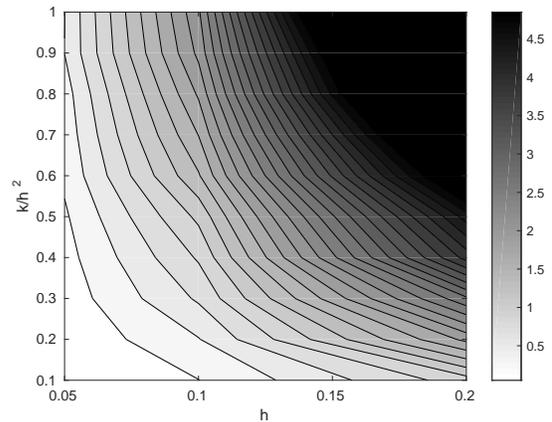, width=8cm}
	\captionof{figure}{Contour plot of the $l_2$-error for second-order scheme.} 
	\label{fig:stest2}
	\end{figure}
\end{minipage}

\subsection{Feller Condition}

\noindent To further test the robustness of the new HOC scheme, we examine the convergence rates achieved when the Feller condition, $2\kappa\theta \geq v^2$, is 
not satisifed for the Cox-Ingersol-Ross (CIR) volatility process \cite{CoxIngRoss}.

We use the default parameters defined in Table~\ref{t:params}, with
exceptions for long-run variance mean $\theta$ and volatility of volatility $v$, which we alter to test the condition as shown in Table~\ref{t:fellerparams}. 

\begin{table}[h]
\centering
\begin{tabular} { c c c }
\toprule
$\theta$ & v & Condition \\
\midrule 
\multirow{3}{*}{0.04}  & 0.7 & $ 2 \kappa \theta < v^2 $ \\
& 0.4 & $ 2 \kappa \theta = v^2 $ \\
& 0.1 & $ 2 \kappa \theta > v^2$ \\
\bottomrule
\end{tabular} \\
\vspace{2mm}
\captionof{table}{Parameters for different regimes of the Feller condition.}
\label{t:fellerparams}
\end{table}

We study the $l_2$ and $l_{\infty}$ -error associated with each
condition. The results are shown in Table~\ref{t:fellerresults}, the $l_2$-error
numerical convergence rates, obtained from a least squares fitted line
as explained earlier,  are $4.0$, $3.9$ and $3.9$ for $v= 0.7$, $0.4 $
and $0.1$, respectively. As a consequence we can confirm the new HOC scheme performs well irrespective of the validity of the Feller condition.   

\begin{table}[h]
\centering
\begin{tabular} { c c c c }
\toprule
Condition & $h$ & $l_2$-error & $l_{\infty}$-error  \\
\midrule
\multirow{3}{*}{$ 2 \kappa \theta < v^2 $} 
& $h=0.2$ & 2.3342 & 0.1930 \\
& $h=0.1$ & 0.0473 & 0.0057 \\
& $h=0.05$ & $ 0.0096 $ & 0.0011 \\
\midrule
\multirow{3}{*}{$ 2 \kappa \theta = v^2 $} 
& $h=0.2$ & 1.3593 & 0.1429 \\
& $h=0.1$ & 0.0289 & 0.0052 \\
& $h=0.05$ & 0.0057 & 0.0010 \\
\midrule
\multirow{3}{*}{$ 2 \kappa \theta > v^2$} 
& $h=0.2$ & 0.9436 & 0.1906 \\
& $h=0.1$ & 0.0394 & 0.0123 \\
& $h=0.05$ & 0.0043 & 9.05 $\cdot 10^{-4}$ \\
\bottomrule
\end{tabular} \\
\vspace{2mm}
\captionof{table}{Numerical convergence results for HOC with varying Feller
  condition.}
\label{t:fellerresults}
\end{table}

\subsection{Hedging performance}

\begin{figure}
\epsfig{file=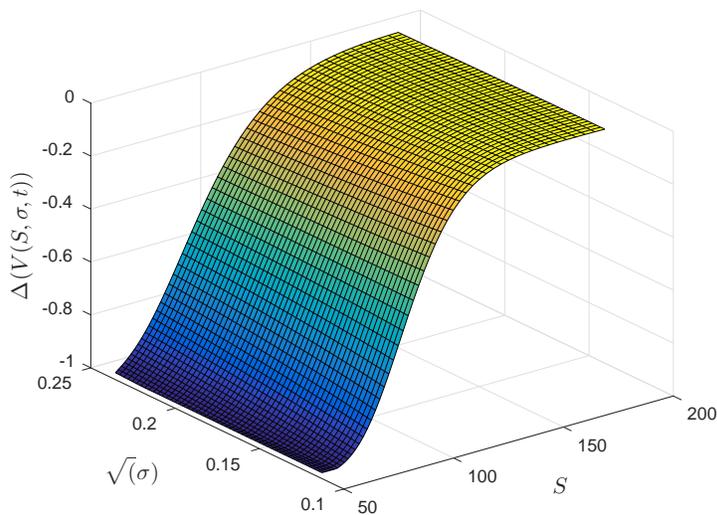, width=10cm}
\caption{ Delta of option with default parameters.} 
\label{f:delta}
\centering
\end{figure}

\noindent The so-called Greeks (partial derivatives of the option price with respect to
independent variables or parameters) are quantities which represent
the market sensitivities of options. Delta measures the sensitivity of
the option price with respect to the price the underlying asset, i.e.\ 
\[ \Delta = \frac{\partial{V}}{\partial{S}}. \]

Delta hedging is a common strategy employed by options traders, an options strategy that aims to hedge the risk associated with price movements in the underlying asset, by offsetting long and short positions. This strategy allows a trader to profit from potential shifts in volatility or the option duration, however to be fully hedged a trader must adapt their portfolio by managing the position in the underlying. In this instance the higher order convergence of our scheme may be of use to traders. 

We propose that the higher-order convergence achieved in the option price will also be represented in the Delta of the option, and as a consequence we will achieve a better hedge.

We calculate the Delta from the option price $V_{i,j}^n\approx V(S_i,\sigma_j,t_n)$. To maintain the order of the scheme we use the following fourth-order approximation formula with the boundaries trimmed to remove the need for extrapolation,
\[ \Delta_{i,j}^n = \frac{1}{S_i} \frac{ V_{i,j-2}^n -8V_{i,j-1}^n +8V_{i,j+1}^n -V_{i,j+2}^n
      }{12h} .\]
Figure~\ref{f:delta} shows the resulting Delta of a European put
option.
Through the same numerical convergence method used for the option
price we examine the convergence of the Delta with respect to a
numerical reference solution. The results are seen in Figures~\ref{fig:test1} and \ref{fig:test2}.
We observe also here that the numerical convergence order agree well with the
theoretical order of the schemes, with the new high-order compact
scheme achieving convergence rates between three and four.

\begin{minipage} {0.5\textwidth}
\begin{figure}[H]
	\centering
	\epsfig{file=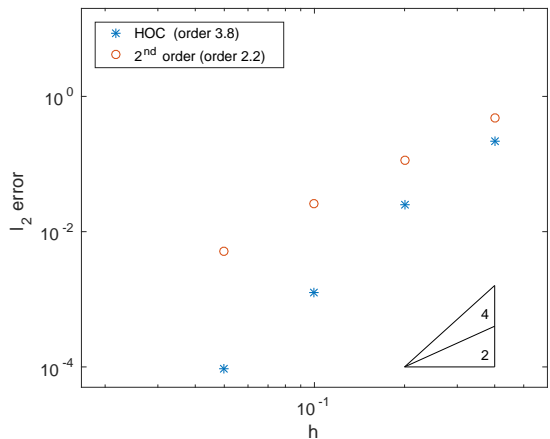, width=8cm}
	\captionof{figure}{$l_2$-error in Delta. } 
	\label{fig:test1}
	\end{figure}
\end{minipage}%
\begin{minipage} {0.5\textwidth}
\begin{figure}[H]
	\centering
	\epsfig{file=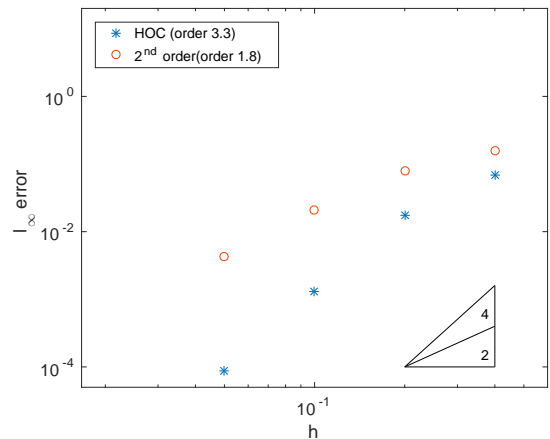, width=8cm}
	\captionof{figure}{ $l_\infty$-error in Delta. } 
	\label{fig:test2}
	\end{figure}
\end{minipage}
%
%
%

\subsection{Delta hedging --- Delta-neutral portfolio}

\noindent We construct a Delta-neutral portfolio $\Pi = P - \Delta S  $ to measure the accuracy of the hedge, the value of this portfolio should not be affected by any change in the underlying asset. We conduct the test on a fine reference grid with mesh-size $h_{\text{ref}}=0.025,$ then we compare the performance of each subsequent mesh-size. For comparative purposes this test is also conducted using the second-order scheme central difference scheme.

We now examine the percentage error introduced into the value of each
portfolio in comparison to the reference grid. This test is conducted
by moving the asset price up or down by a fixed amount. The results
for this experiment are shown in Table~\ref{t:deltadown} and Table~\ref{t:deltaup}, with the
parameters given in Table~\ref{t:params}. We observe that the high-order scheme offers a better delta hedge, even on a
coarser grid.

\begin{table}[h]
\centering
\begin{tabular} { c c c }
\toprule
Mesh Size & HOC & 2nd order  \\
\midrule
$h=0.4$ & 5.7649 & 10.3354  \\
$h=0.2$ & 0.3505 & 2.2765 \\
$h=0.1$ & 0.0083 & 0.5598 \\
$h=0.05$ & $7.33\cdot{10^{-4}} $ & 0.1137 \\
\bottomrule
\end{tabular}
\captionof{table}{Percentage error in portfolio value for a move down
  in the underlying.}
\label{t:deltadown}
\end{table}

\begin{table}[h]
\centering
\vspace{2mm}
\begin{tabular} { c c c }
\toprule
Mesh Size & HOC & 2nd order \\
\midrule
$h=0.4$ & 4.6914 & 6.4067    \\
$h=0.2$ & 0.2980 & 1.0895  \\
$h=0.1$ & 0.0074 & 0.2417  \\
$h=0.05$ & ${7.86}\cdot{10^{-4}}$ & 0.0493 \\
\bottomrule
\end{tabular}
\captionof{table}{Percentage error in portfolio value for a move up in
  the underlying.}
\label{t:deltaup}
\end{table}

\section{Conclusions}
\label{sec:conc}

\noindent We have derived a new high-order compact finite difference method for
option pricing in stochastic volatility jump models. Numerical
experiments confirm high-order convergence in both the option price
and the Delta of the option. The method is based on an
implicit-explicit scheme in combination with high-order compact finite
difference stencils for solving the partial integro-differential
equation. It can be implemented in a highly efficient manner and can
be used to upgrade existing finite difference codes. Compared to
finite element
methods, it is very parsimonious in terms of memory requirements
and computational effort, since it achieves high-order convergence
without requiring additional unknowns (unlike finite elements with higher
polynomial order).
 Examples of a Delta hedged portfolio provide clear evidence that the
 high-order scheme is valuable for industry professionals seeking to
 calculate the relevant Delta accurately and requiring fastest
 computational time.

The American option pricing problem which requires solving a free
boundary problem involving the partial integro-differential equation
\eqref{eq:Bates} can in principle be approached by combining the
high-order compact scheme
presented in this paper with standard methods like projected
successive overrelaxation (PSOR) or penalty methods.
The key challenge, however, will be to retain high-order convergence of the scheme in view of
limited regularity across the free boundary. 

A straightforward extension of this paper is the introduction of the
so-called 
SVCJ model which allows for jumps in both returns and volatility.
As a second extension, one can combine the method presented in
this paper with high-order alternating direction implicit methods \cite{DuMi17} and
with sparse grids methods \cite{HeHeEhGu17,DuHeMi17}.
We leave these extensions for future research.

\bigskip\noindent{{\bf Acknowledgements.}\newline
BD acknowledges partial support by the Leverhulme Trust research project grant `Novel discretisations for higher-order nonlinear PDE' (RPG-2015-69). 
AP has been supported by a studentship under the EPSRC Doctoral
Training Partnership (DTP) scheme (grant number EP/M506667/1).
The authors are grateful to the anonymous referees for helpful
remarks and suggestions.
}

\end{document}